\documentclass[twocolumn,showpacs,preprintnumbers,amsmath,amssymb]{revtex4}

\usepackage{graphicx}
\usepackage{dcolumn}
\usepackage{bm}

\newcommand{\be}{\begin{equation}}  
\newcommand{\ee}{\end{equation}}  
\newcommand{\bear}{\begin{eqnarray}}  
\newcommand{\eear}{\end{eqnarray}}  
\newcommand{\ba}{\begin{array}}  
\newcommand{\ea}{\end{array}}

\newskip\humongous \humongous=0pt plus 1000pt minus 1000pt

\newif\ifdtup

\def\oldreffmt#1{\rlap{[#1]} \hbox to 2\parindent{}}

\def\figfmt#1{\rlap{Figure {#1}} \hbox to 1in{}}  
  
\def\ie{\hbox{\it i.e.}{}}	  
\def\eg{\hbox{\it e.g.}{}}

\def\Tr{\mathop{\rm Tr}}

\def\ket#1{\left| #1\right\rangle}  
\def\VEV#1{\left\langle #1\right\rangle}

\def\slash#1{#1\!\!\!/\!\,\,}  
\def\beq{\begin{equation}}  
\def\eeq{\end{equation}}  
\def\bea{\begin{eqnarray}}  
\def\eea{\end{eqnarray}}  
\def\half{\frac{1}{2}}  
  
\def\bq{\begin{quote}}  
\def\eq{\end{quote}}

\def\half{\frac{1}{2}}       

\relax  

\newdimen\tdim  
\tdim=\unitlength  
\def\bar{\overline}

\begin{document}

\preprint{FERMILAB-Pub-03/071-T}
\title{Chiral Multiplets of Heavy-Light Mesons}

\author{William A. Bardeen, Estia J. Eichten, Christopher T. Hill}

\email{bardeen@fnal.gov,eichten@fnal.gov,hill@fnal.gov}

\affiliation{
 {{Fermi National Accelerator Laboratory}}\\
{{\it P.O. Box 500, Batavia, Illinois 60510, USA}}
}%

\date{\today}

\begin{abstract}
The recent discovery of a narrow resonance in $D_s\pi^0$
by the BABAR collaboration is consistent with
the interpretation of a heavy $J^P$ $(0^+,1^+)$ 
spin multiplet. This system is the
parity partner of the groundstate $(0^-,1^-)$
multiplet, which we argue is required in
the implementation of $SU(3)_L\times SU(3)_R$ chiral symmetry  
in heavy--light meson systems.  
The  $(0^+,1^+) \rightarrow (0^-,1^-)+\pi$  transition couplings 
satisfy a Goldberger-Treiman
relation, $g_\pi = \Delta M/f_\pi$, where $\Delta M$
is the mass gap. 
The BABAR resonance fits the $0^+$ state, with a kinematically
blocked principal decay mode to $D + K$.
The allowed $D_s+\pi$, $D_s+2\pi$, and electromagnetic
transitions are computed from the full chiral
theory and found to be suppressed, consistent with the narrowness
of the state.  This state establishes
the chiral mass difference for all such heavy-quark chiral
multiplets, and precise predictions exist for the 
analogous $B_s$ and strange doubly-heavy baryon states.
\end{abstract}

\maketitle

\section{Introduction}

Recently the BABAR collaboration has
reported the observation of
a narrow resonance  in $D^+_s\pi^0$ with a mass of $2317$ MeV \cite{BABAR}.  
There is also a hint
of a second state in  $D_s\pi^0\gamma$ with a mass $2460$ MeV.
The mass difference between the $D^*_s(2317)$ and the 
well established lightest charm-strange meson,
$D_s$, is $\Delta M = 349$ MeV. This is less than the kaon mass, 
thus kinematically forbidding the
decay $D_s^*(2317) \rightarrow D_{u,d}+K$. 
In the present paper we will argue that both of these states 
are indeed members the heavy  $(0^+,1^+)$ spin multiplet
usually identified with the $ j_\ell = 1/2$,
$p$-wave valence light--quark system.  

Heavy-light systems involving a single valence light--quark, heavy-light
mesons and baryons with two heavy--quarks, can be viewed as a ``tethered"
constituent--quark. 
In QCD the light--quark chiral symmetry is
spontaneously broken and the symmetry is realized nonlinearly,
usually described via chiral Lagrangians with nearly massless pions.
Suppose we could somehow modify QCD to restore both the explicit
and the spontaneously broken light--quark chiral symmetries
while maintaining the confining properties of QCD. For example,
a hypothetical four--fermion interaction involving only the light--quarks 
could trigger such a phase change.

In the symmetric phase, the constituent--quark would be expected
to become massless -- preserving the chiral symmetry. However,
a tethered constituent--quark cannot become massless, due 
to the presence of the heavy--quark(s) to which it is tethered. 
In this case, 
the confined heavy-light hadrons {\em would be forced} to appear 
in parity--doubled bound states transforming as linear representations
of the light--quark chiral symmetry. The heavy--light hadrons 
would be described by an effective field theory of these parity
doubled states. The light hadron states would also appear in 
either massless chiral representations, like the constituent 
quark, or in massive parity--doubled representations. The pions would
be forced into a linear chiral multiplet with 
scalar mesons. 

As we adiabatically turn off the 
interactions modifying QCD and restore the 
spontaneously-broken chiral phase, we would expect the 
effective field theory Lagrangians  describing the light hadrons, and 
the heavy-light hadrons, to evolve smoothly, with the main effect 
being the shift in the scalar mass terms (this is
irrespective of the order of the associated chiral phase transition;
it is simply the flow of the Lagrangian parameters that
matters to us). Linear $\Sigma$-models 
have been used successfully to describe the physics of light 
hadrons since their introduction by Gell-Mann and L\'{e}vy \cite{MGM} in 
1960. For heavy--light systems, the effective field theory would 
require parity doubled representations of the hadrons with the 
doubling degeneracies being lifted through couplings to the light 
quark chiral fields. An essential feature of this dynamical symmetry 
breaking mechanism is the Goldberger-Treiman relation 
\cite{Goldberger}
between the 
mass--splitting in chiral multiplets and the couplings to soft pions. 
We will use this analogy to construct a $\Sigma$--model for the tethered
constituent--quark states that invokes linear realizations of the 
light--quark chiral symmetry and a smooth interpolation to the chiral 
broken phase. Effective Lagrangians with both heavy--quark symmetry
and linearly realized chiral symmetry have been constructed previously
to generate a simplified dynamical model of the bound meson states 
of heavy and light--quarks \cite{Bardeen,nowak,eeg}.

The main consequence is that this newly observed multiplet is, to a good
approximation, the {\em chiral partner} of the $(0^-,1^-)$
groundstate.  Physically, this means that the
two orthogonal linear combinations of meson fields, 
$D(0^+,1^+)+ D(0^-,1^-)$
and $D(0^+,1^+)- D(0^-,1^-)$,
have well defined transformation properties under 
$SU(3)_L\times SU(3)_R$, transforming as 
(approximately) pure $(1,3)$ and
$(3,1)$ respectively. 
The parity doubling implies that
the  main decay transitions $(0^+,1^+)\rightarrow (0^-,1^-) +``\pi"$,
where $``\pi"$ refers to any of the pseudsoscalar octet mesons,
are governed by a 
Goldberger--Treiman (GT) relation, 
$g_\pi = \Delta M/f_\pi$, where $\Delta M$ is
the $0^-$-- $0^+$ mass difference,
$g_\pi$ is the $0^+\rightarrow 0^-\pi$ coupling
constant, and $f_\pi$ the pion decay
constant. $\Delta M$ represents a left-right transition,
the analogue of the mass of the nucleon, which occurs in
the successful GT relation $g_{NN\pi} =m_N/f_\pi$.
The observed magnitude of $\Delta M$ has a simple physical
explanation: the heavy-light meson contains 
a single constituent--quark, while
the nucleon contains three, and we therefore
expect that $g_{\pi}\approx g_{NN\pi}/3$,
hence $\Delta M \approx m_N/3$. The $D_s(2317)$ is the 
first clear observation
of this more general phenomenon, which we
expect to hold in
all heavy-light mesons, doubly-heavy
baryons, and yields correspondingly narrow
states in the $B_s$-mesons, and 
the strange heavy-heavy-light baryons, 
$ccs$, $cbs$ and $bbs$.

The GT relation implies
that the expected rates for the nonstrange resonances
to decay through pionic transitions are now
determined precisely for all of the analogue
systems,  and we tabulate them. 
The BABAR resonances, however, interpreted as the $0^+$ and $1^+$
states, would have had principal GT transitions, 
decaying through a kaonic
transition, $D_s(2317) \rightarrow D_{u,d}+K$, but this is blocked 
by kinematics. It must therefore proceed through $SU(3)$
breaking effects, decaying by 
$D_s(2317) \rightarrow D_{u,d}+(\eta\rightarrow \pi^0)$, emitting
a virtual $\eta$ that then mixes with
the $\pi^0$ through isospin violating
effects. We compute its width 
and find it is indeed narrow. We also tabulate
the widths for all analogue processes.  
We further show that electromagnetic transitions are 
indeed, and somewhat remarkably,
suppressed, since they involve cancellations between
the heavy and light magnetic moments. Again,
we tabulate  rates for the analogue systems. 
The overall picture of the chiral
structure of the heavy-light systems works
quite well.  

In the next section we begin with the
familiar fact that heavy-light mesons $H\sim \bar{Q}q$,  
containing one heavy--quark $Q$ and
one light--quark $q$, are subject to powerful
symmetry constraints. The heavy--quark symmetry must apply in the limit 
$m_Q\rightarrow \infty$ \cite{Eichten,Isgur,Georgi}, 
typically implying vanishing hyperfine splitting
effects, and leading to degenerate 
spin multiplets, as in the $(0^-,1^-)$ groundstate of
the system.   In addition the light--quark chiral symmetries
of QCD must apply. Hence, in the limit $m_q\rightarrow 0$, where $q=(u,d,s)$,
any Lagrangian must be invariant under the $SU(3)_L\times SU(3)_R$
chiral symmetry, broken by the light--quark mass matrix and
electromagnetism.   Together these heavy--quark (HQ)
and chiral light--quark (LQ) symmetries control the interactions
of heavy--light (HL) mesons with pions and $K$--mesons, etc.
Presently we will not delve into chiral-constituent--quark models.
Rather, we
write directly the chiral Lagrangian for the two heavy-quark
multiplets, $H\sim (0^-,1^-)$, and $H' \sim(0^+,1^+)$, implementing
both HQ symmetry and chiral $SU(3)_L\times SU(3)_R$.

\noindent

\section{Effective Lagrangians}

We begin in the limit
in which the linear chiral symmetry is an
exact symmetry of the vacuum.
In this limit the heavy--light $(0^-,1^-)$ multiplet is 
degenerate with the $(0^+,1^+)$ multiplet. 
We must therefore
introduce four independent heavy--meson
fields, $H\; ( H')$ are $0^-$
($0^+$) scalars, while $H_\mu\; (H'_\mu)$ are
$1^-$ ($1^+$) vectors. 
Heavy--quark symmetry is implemented by constructing
multiplets for a fixed four velocity supersector, $v_\mu$.
One heavy--spin  multiplet
consists  of the $0^+$ scalar together with
the abnormal parity ($1^+$) vector as $(H', H'{}^{\mu})$.
Under heavy spin $O(4)=SU(2)_h\times SU(2)_l$
rotations the $(H,' H'{}^{\mu})$ mix analogously
to $(H, H^\mu)$, transforming as the ${\bf 4}$
representation of $O(4)$ 
(the four--velocity label $_v$, and $SU(3)$ $^i$
indices are understood):
\be
{\cal{H}}'  =  
(i\gamma^5 H' + \gamma_\mu H'{}^{\mu})\left(\frac{1+\slash{v}}{2}\right)
\ee
The other multiplet
consists of the usual $0^-$ scalar and a $1^-$ vector $(H, H^{\mu})$:
\be
{\cal{H}}   =  
(i\gamma^5 H + \gamma_\mu H^{\mu})\left(\frac{1+\slash{v}}{2}\right)
\ee
 Note that we have the constraint $v_\mu H^\mu = 0$.
We  have introduced the caligraphic ${\cal{H}}$ and
${\cal{H}}'$ with the explicit projection factors.
We have reversed the order of 
the heavy spin projection and
the field components because it is more convenient for
writing manifestly chirally inivariant operators
this way. 
The field ${\cal{H}}'$ has overall
odd parity, while ${\cal{H}}$ is  even.
With either of
these fields a properly normalized kinetic term
can be written as:
\be
-i\half \Tr(\overline{{\cal{H}}} v\cdot\partial
{\cal{H}} )
\ee
where the trace extends over Dirac and flavor indices
(see Appendix A for the full normalization conventions).

To implement a linear
chiral symmetry multiplet structure in the HL sector
we construct left-handed and right-handed linear 
combinations of the heavy spin multiplets.  
We define the following
chiral combinations:
\bea
& & {\cal{H}}_{L} = \frac{1}{\sqrt{2}}\left({\cal{H}} - i {\cal{H}}'\right)  \qquad
{\cal{H}}_{R} = \frac{1}{\sqrt{2}}\left({\cal{H}} + i {\cal{H}}'\right)
\eea
Under $SU(3)_L\times SU(3)_R$ these fields transform as 
${\cal{H}}_R\sim (1,3)$ and ${\cal{H}}_L\sim (3,1)$ respectively.

To describe the light mesons
we introduce the chiral field $\Sigma$, 
transforming as $(\bar{3},3)$ under $SU(3)_L\times SU(3)_R$.
The usual linear $\Sigma$-model Lagrangian is:
\bea
{\cal{L}}_L & = &  \frac{1}{4}\Tr(\partial_\mu\Sigma^\dagger\partial^\mu\Sigma) 
+ 
\kappa\Tr ({\cal{M}}_q\Sigma + h.c.)- V(\Sigma)
\nonumber \\
\eea
where ${\cal{M}}_q$ is the light--quark mass matrix,
representing explicit $SU(3)_L\times SU(3)_R$ breaking,
and:
\bea
V(\Sigma) & = &
 -\frac{1}{4}\mu_0^2\Tr(\Sigma^\dagger\Sigma) 
+ \frac{1}{8} \lambda_0 \Tr(\Sigma^\dagger\Sigma\Sigma^\dagger\Sigma) 
\nonumber \\
& & 
- \Lambda_0(e^{i\theta}\det\Sigma + h.c.) +...
\eea
where we have included $U(1)_A$ breaking effects
through a `t Hooft determinant term.
The field $\Sigma$ is a $3\times 3$ complex matrix.
The imaginary components of  $\Sigma$ are the $0^-$ nonet,
including $\pi, K, \eta,\eta'$,
while the real components form a $0^+$ nonet. 

In the
chiral symmetric phase, $\VEV{\Sigma}=0$,
and the $0^+$ and $0^-$ octets
are degenerate, forming a massive parity doubled
nonet.
When the chiral symmetry is spontaneously broken,
 $\VEV{\Sigma} = I_3f_\pi$ and
the $0^+$ nonet becomes heavy,  while
the $0^-$ octet becomes a set of
 massless Goldstone bosons, the $\eta'$ receiving
a nonzero mass from the `t Hooft determinant.

In the broken 
symmetry phase we can write:
\be
\Sigma = \xi \tilde{\sigma} \xi
\qquad \xi = \exp(i\pi\cdot\lambda/2f_\pi)
\ee
where the $0^+$ nonet field is:
\be
\tilde{\sigma} = \sqrt{\frac{2}{3}}\sigma I_3 + \sigma^a\lambda^a
\ee
and $\VEV{\sigma}= \sqrt{3/2}f_\pi$.
If we then take 
the $0^+$ nonet mass to infinity, holding $f_\pi$
fixed, we can decribe
the octet of pseudoscalar mesons in the 
nonlinear $\Sigma$-model:
\beq
\Sigma = f_\pi
\exp(i\pi^a\cdot\lambda^a/f_\pi)  
\eeq   
Note that $f_\pi = 93.3$ MeV
in this normalization. 
The nonlinear chiral Lagrangian takes the form:
\bea
{\cal{L}}_L & = & \frac{1}{4}\Tr(\partial_\mu\Sigma^\dagger\partial^\mu\Sigma) 
+  \half\kappa\Tr ({\cal{M}}_q\Sigma + h.c.)
\eea
where  
$\kappa=f_\pi m_\pi^2/(m_u+m_d)$
fits the meson masses, yielding the Gell-Mann--Okubo
formula. The expansion of the mass term 
in the meson fields to quadratic order yields
an isospin violating  $\pi^0$ $\eta$ mixing term that
we will require later:
\bea
\label{choeq}
{\cal{L}}_L & = & ... \;
+ \frac{m_\pi^2(m_u-m_d)}{\sqrt{3}(m_u+m_d)}\pi^0 \eta
\eea

We now write an effective Lagrangian involving
both the HL mesons and the $\Sigma$ field,
implementing HQ symmetry and chiral symmetry.
The lowest order effective Lagrangian  to first
order
in an expansion in the chiral field $\Sigma$, 
and to zeroth order
in $(1/m_Q)$ is \cite{Bardeen}:
\bea
\label{chiral1}
{\cal{L}}_{LH} & = & -i\half \Tr(\overline{{\cal{H}}}_L v\cdot\partial
{\cal{H}}_{L} )
 -i\half \Tr(\overline{{\cal{H}}}_{R}  v\cdot\partial  {\cal{H}}_{R} )
\nonumber \\
& & -\frac{{g_{\pi}} }{4}
\left[  \Tr(\overline{{\cal{H}}_{L}}\Sigma^\dagger
 {\cal{H}}_{R}) +\Tr(\overline{{\cal{H}}_{R}}\Sigma
 {\cal{H}}_{L}) 
\right]
\nonumber \\
& & -\Delta_{} 
\left( \Tr(\overline{{\cal{H}}_{L}} {\cal{H}}_{L}) 
+\Tr(\overline{{\cal{H}}_{R}} {\cal{H}}_{R}) 
\right)
\nonumber \\
&  & \!\!\!\!\!\!\!\!\!\!\!\!\!\!\!
+i\frac{g_{A}}{ 2f_\pi}  
\left[  \Tr(\overline{{\cal{H}}_{L}}\gamma^5(\slash{\partial}\Sigma^\dagger)
 {\cal{H}}_{R}) 
 -\Tr(\overline{{\cal{H}}_{R}}\gamma^5(\slash{\partial}
 \Sigma)
 {\cal{H}}_{L})
\right]
\nonumber \\
& & + \;\; ...
\eea
The $\Delta$ term can be ``gauged away'' by a
reparameterization transformation on the fields, so we 
henceforth drop it.

Terms can be added at first order in
$(1/m_Q)$ to accomodate the intramultiplet
hyperfine mass splitting effects:
\bea
\label{hyperfine}
{\cal{L}}_{0,hyperfine} 
& = & 
\nonumber \\
& &
\!\!\!\!\!\!\!\!\!\!\!\!\!\!\!\!\!\!\!\!\!\!\!\!\!\!\!\!\!\!
\frac{\Lambda_{QCD}^2}{12m_Q}
\left[ k_1 \Tr(\overline{{\cal{H}}}_L\sigma_{\mu\nu}
{\cal{H}}_L\sigma^{\mu\nu}) +
 k_2\Tr(\overline{{\cal{H}}}_R\sigma_{\mu\nu}
{\cal{H}}_R\sigma^{\mu\nu})   
\right]
\nonumber \\
\eea
Parity symmetry implies invariance under $L\leftrightarrow R$,
and $\Sigma\leftrightarrow \Sigma^\dagger$ hence:
\be
k \equiv k_1 = k_2
\ee
There are additional terms of order $1/m_Q$, such
as $\Tr(\overline{{\cal{H}}}_L(v\cdot\partial)^2{\cal{H}}_L)
+ (L\leftrightarrow R)$. 

The hyperfine splitting effects to first order
in $\Sigma$ and first order in $1/m_Q$ are 
$LR$ transition terms of the form:
\bea
\label{hyperfine}
{\cal{L}}_{1,hyperfine} 
& = &
\frac{k' \Lambda_{QCD}^2}{12 m_Q f_\pi}
\left[ \Tr(\overline{{\cal{H}}}_L\sigma_{\mu\nu}
 \Sigma^\dagger {\cal{H}}_R\sigma^{\mu\nu}) + h.c.
\right]
\nonumber \\
\eea
Since these terms are overall second order effects 
we expect them to be small $k' << k$.

We can perform redefinitions
of the heavy fields to bring them into linear flavor
$SU(3)$ representations in the parity eigenbasis:
\beq
 {\cal{H}}_L = \frac{1}{\sqrt{2}} \xi^\dagger ({\cal{H}} -i {\cal{H}}')
\qquad
 {\cal{H}}_R = \frac{1}{\sqrt{2}} \xi ({\cal{H}} + {i\cal{H}}') 
\eeq
and the Lagrangian now takes the form:
\bea
\label{chiral2}
{\cal{L}}_{LH} & = & -\half \Tr(\overline{ {\cal{H}}}
v\cdot (i{\partial}
+ {{\cal{V}}}) {\cal{H}} )
-\half \Tr(\overline{ {\cal{H}}}{}'v\cdot(i{\partial}+ {{\cal{V}}})
 {\cal{H}}{}' )
\nonumber \\
& & +i\half G_A\Tr(\overline{ {\cal{H}}}{}'
v\cdot{{\cal{A}}} {\cal{H}} )
-i\half G_A\Tr(\overline{ {\cal{H}}}
v\cdot{{\cal{A}}} {\cal{H}}{}' )
\nonumber \\
& & +\frac{{g_\pi } }{4}
\left[  \Tr(\overline{ {\cal{H}}}{}'\tilde{\sigma}
  {\cal{H}}{}') -\Tr(\overline{ {\cal{H}}}
\tilde{\sigma}  {\cal{H}})   
\right]
\nonumber \\
&  & \!\!\!\!\!\!\!\!\!\!\!\!\!\!\!\!\!\!\!\!
+\frac{g_A}{2f_\pi} 
\left[  \Tr(\overline{ {\cal{H}}}'\gamma^5\gamma_\mu 
\{ {\cal{A}}^\mu,\tilde{\sigma} \}
  {\cal{H}}') 
 -\Tr(\overline{ {\cal{H}}}{}\gamma^5\gamma_\mu 
 \{ {\cal{A}}^\mu,\tilde{\sigma} \}
 {\cal{H}})
\right]
\nonumber \\
&  & 
+\frac{g_A}{2f_\pi} \Tr(\overline{ {\cal{H}}}'\gamma^5\gamma_\mu 
(\partial^\mu\tilde{\sigma}-i[{\cal{V}}^\mu,\tilde{\sigma}])
  {\cal{H}}) 
 \nonumber \\
&  & 
 +\frac{g_A}{2f_\pi} \Tr(\overline{ {\cal{H}}}{}\gamma^5\gamma_\mu 
 (\partial^\mu\tilde{\sigma}-i[{\cal{V}}^\mu,\tilde{\sigma}])
 {\cal{H}}')
  \nonumber \\
&  & +\;\;\; ...
\eea
where:
\be
{\cal{V}}_\mu  = \half(\xi^\dagger \partial_\mu \xi 
+ \xi  \partial_\mu \xi^\dagger)
= \frac{i}{8f^2_\pi}[\tilde{\pi}, \partial_\mu \tilde{\pi}] + ...
\ee
\be
{\cal{A}}_\mu  = i\half( \xi^\dagger \partial_\mu \xi 
- \xi  \partial_\mu \xi^\dagger)
= -\frac{1}{2f_\pi}\partial_\mu \tilde{\pi} + ...
\eeq
where $\tilde{\pi}=\sqrt{2/3}\;\eta' + \pi^a\lambda^a$.
We have introduced a phenomenological
parameter $G_A$, that
is unity in lowest order model of eq.(\ref{chiral1}), but
that can, in principle, receive corrections.

In the chiral symmetric phase the multiplets
$ {\cal{H}}{}'$ and $ {\cal{H}}{}$ are degenerate in mass.
In the broken phase, however, we have $\VEV{\tilde{\sigma}}=f_\pi I_3$, and
from eq.(\ref{chiral2}) we learn that the physical
mass of the $ {\cal{H}}{}'$ state
is elevated by the amount $+g_\pi f_\pi/2$, while
the $ {\cal{H}}{}$ state is depressed by $-g_\pi f_\pi/2$
(see the Appendix for normalization conventions).
The Goldberger-Treiman relation is therefore obtained
relating the mass difference $\Delta M$ to the
coupling constant $g_\pi$:
\be
\Delta M = g_\pi f_\pi
\ee
$\Delta M$ is the mass difference between the multiplets
that is now measured to be  $349$ MeV from
the BABAR results. This implies $g_\pi = 3.73$.

Note that we can decouple the heavier field $ {\cal{H}}'$
and the $0^+$ nonet,
yielding an effective chiral
Lagrangian for the lower energy
field $ {\cal{H}}$ alone:
\bea
{\cal{L}}_{LH} & = & -\half \Tr(\overline{ {\cal{H}}}
v\cdot (i{\partial}
+ {{\cal{V}}}){ {\cal{H}}} )
-g_A 
\Tr \overline{  {\cal{H}}}\gamma^5 \slash{{\cal{A}}}
 {\cal{H}}
\nonumber \\
\eea
This Lagrangian 
\cite{wise} contains relatively limited information,
compared to eq.(\ref{chiral2}),
and will not apply on energy scales
approaching $\Delta M$ above the groundstate
mass. 

The hyperfine mass splitting effects
now take the form:
\bea
\label{hyperfine}
{\cal{L}}_{LH,hyperfine} 
& = &\nonumber \\
& &\!\!\!\!\!\!\!\!\!\!\!\!\!\!\!\!\!\!\!\!\!\!\!\!\!\!\!\!\!\!
k \frac{\Lambda_{QCD}^2}{12m_Q}
\left[\Tr(\overline{ {\cal{H}}}{}'\sigma_{\mu\nu}
  {\cal{H}}{}'\sigma^{\mu\nu}) +
 \Tr(\overline{ {\cal{H}}}\sigma_{\mu\nu}
  {\cal{H}}\sigma^{\mu\nu})   
\right]
\nonumber \\
\eea
We have assumed $k' << k$ is negligible, as
per the discussion of ordering the strengths
of various terms. This implies that the hyperfine
splitting within the heavy $(0^+,1^+)$
multiplet is identical to
that in the groundstate $(0^-,1^-)$ mesons.

\section{Spectrum}

From the Lagrangian of
eq.(\ref{chiral2})
we see that the 
chiral multiplet structure, together with HQ
symmetry controls, the masses within
the $(0^+, 1^+)$ multiplet.
The spin-weighted center
of mass  of any $(0^+, 1^+)$ multiplet 
will have a universal $\Delta M(m_Q)$
above the corresponding spin-weighted
groundstate in all heavy--light systems.
This is weakly dependent upon $m_Q$,
and approaches  a universal value $\Delta M(\infty)$
in the heavy--quark symmetry limit
limit, $m_Q\rightarrow \infty$.

The observed $D_s(0^+)$ resonance
in BABAR measures $\Delta M(m_c)$. 
$\Delta M(m_c)$ is therefore determined by the
mass difference of the $D_s(0^+,2317)$ and the
groundstate $D_s(0^-,1969)$ to be:
\be 
\Delta M(m_c)=349\; \makebox{MeV}.
\ee
A predicted value of $\Delta M(\infty)\approx 338$ MeV 
was obtained in  \cite{Bardeen}
from a fit to the HL chiral constituent--quark model.

Using  $\Delta M(m_c)$ we predict the $D_s(1^+)$ mass:
\be
M(D_s(1^+)) = 2460\; \makebox{MeV}
\ee
from the sum of the  $D_s(1^-,2112)$ mass and $\Delta M(m_c)$.
This is in good agreement with the hinted second resonance
in $D_s\pi^0\gamma$ in the BABAR data.

In the nonstrange
$D^\pm(0^+, 1^+)$ and $D^0(0^+, 1^+)$ multiplets
the chiral mass gap is also given by
the measured value $\Delta M(m_c)$.
We therefore predict:
\bea
M(D^\pm(0^+))& = & 2217\; \makebox{MeV}; 
\nonumber \\
M(D^\pm(1^+)) & = & 2358 \; \makebox{MeV};
\nonumber \\
M(D^0(0^+))& = & 2212 \;  \makebox{MeV}; 
\nonumber \\
M(D^0(1^+)) & = & 2355 \;  \makebox{MeV}. 
\eea
There will be corrections of order $\Lambda_{QCD}/m_c$
to the inferred value of the universal
$\Delta M(\infty)$, from the
center of mass.
The $B$ system will provide a better determination
of the heavy-quark symmetry limit and
the chiral mass gap, $\Delta M(\infty)$.
We have no prediction for these corrections at present
so we take $\Delta M(m_b) = \Delta M(m_c)\pm 35$ MeV.
In the B system we therefore predict:
\be
M(B^\pm(0^+))=M(B^0(0^+))=5627\pm 35\; \makebox{MeV}.
\ee
The  $M(B^\pm(1^-))$ and $M(B^0(1^-))$
masses must be inferred from heavy-quark symmetry.
This is an intramultiplet
hyperfine splitting, above the $M(B^\pm(0^+))=5627$ MeV
groundstate.
In the $B$-system it
is reduced by a factor
of $m_c/m_b = 0.33$ relative to the corresponding
$M(D(1^-))-M(D(0^-)) = 142$ MeV. Hence,
we have:
\be
 M(B^\pm(1^-))-M(B^\pm(0^-)) = 47\; \makebox{MeV}.
 \ee
We thus predict: 
\be
M(B^\pm(1^+))=M(B^0(1^+))=5674\pm 35\; \makebox{MeV}
\ee
For the $B_s$ system we have the 
established groundstate mass
of $M(B_s(0^-))=5370$ MeV and we likewise infer:
\be
M(B_s(1^-))=5417\; \makebox{MeV}.
\ee
From this we predict:
\bea
M(B_s(0^+)) & = & 5718\pm 35\; \makebox{MeV}
\nonumber \\
M(B_s(1^+)) & = & 5765\pm 35\; \makebox{MeV}.
\eea

\section{Pionic Transitions}

\vskip .1in
\noindent
{\bf (A) Intermultiplet Transitions}
\vskip .1in
\noindent

The chiral structure of the theory controls the decays of the 
form $(0^+,1^+)\rightarrow (0^-,1^-)+\pi$. These decays 
between multiplets proceed through the axial coupling term:
\be
\label{trans1}
+i\half G_A\Tr(\overline{ {\cal{H}}}{}'
v\cdot{{\cal{A}}} {\cal{H}} )
-i\half G_A\Tr(\overline{ {\cal{H}}}
v\cdot{{\cal{A}}} {\cal{H}}{}' )
\ee
All such transitions for HL mesons and
doubly-heavy baryons are governed by the same 
amplitude, and differ only by phase space.

\vskip .1in
\noindent
{\em (i) $D_{u,d}(0^+,1^+)\rightarrow 
D_{u,d}(0^-,1^-)+\pi$ }
\vskip .1in
\noindent

The amplitudes $D_{u,d}(0^+,1^+)\rightarrow 
D_{u,d}(0^-,1^-)+\pi$ follow from writing eq.(\ref{trans1})
in component form. For example,
the $\pi^0$ transition is:
\be
+i\half G_A\Tr(\overline{ {\cal{H}}}{}'
v\cdot{{\cal{A}}} {\cal{H}} ) \rightarrow
-\frac{iG_A}{2f_\pi}(\overline{D}'_\mu D^\mu- \overline{D}' D)
v_\nu\partial^\nu \pi^0
\ee
leading to the partial width:
\bea
\Gamma(D_{u,d}(0^+,1^+)\rightarrow 
D_{u,d}(0^-,1^-)+\pi^0) & &
\nonumber \\
& & \!\!\!\!\!\!\!\!\!\!\!\!\!\!\!\!\!\!\!\!\!\!\!\!
\!\!\!\!\!\!\!\!\!\!\!\!\!\!\!\!\!\!\!\!\!\!\!\!
\!\!\!\!\!\!\!\!\!\!\!\!\!\!\!\!\!\!\!\!\!\!\!\!
\!\!\!\!\!\!\!\!\!\!\!
= \frac{G_A^2(\Delta M)^2 }{4\pi f_\pi^2}|\vec{p}_\pi|
= 164 \;G_A^2\; \makebox{MeV}.
\eea
The charged pion rate differs by a $\sqrt{2}$ in amplitude:
\be
\Gamma(D_{u,d}(0^+,1^+)\rightarrow 
D_{u,d}(0^-,1^-)+\pi^+) = 326 \;G_A^2\; \makebox{MeV}.
\ee
Thus the partial widths of the $D_{u,d}(0^+)$ and 
$D_{u,d}(1^+)$ are identical, and the total widths
are $490 \; G_A^2$ MeV.  We expect $G_A\approx 1$.

\vskip .1in
\noindent
{\em (ii) $B_{u,d}(0^+,1^+)\rightarrow 
B_{u,d}(0^-,1^-)+\pi$ } 	      
\noindent
\vskip .1in
We expect identical results for the $B_{u,d}(0^+,1^+)\rightarrow 
B_{u,d}(0^-,1^-)+\pi$ transitions:
\bea
\Gamma(B_{u,d}(0^+,1^+)\rightarrow 
B_{u,d}(0^-,1^-)+\pi^0) & &
\nonumber \\
& & 
\!\!\!\!\!\!\!\!\!\!\!\!\!\!\!\!\!\!\!\!\!\!\!\!
\!\!\!\!\!\!\!\!\!\!\!
= 164 \;G_A^2\; \makebox{MeV}.
\eea
\be
\Gamma(B_{u,d}(0^+,1^+)\rightarrow 
B_{u,d}(0^-,1^-)+\pi^+) = 326 \;G_A^2 \; \makebox{MeV}.
\ee

\vskip .1in
\noindent			      
{\em (iii) $D_{s}(0^+,1^+)\rightarrow 			      
D_{s}(0^-,1^-)+\pi^0$ }
\vskip .1in
\noindent
The decay proceeds by the emission of a virtual
$\eta$ that then mixes to the $\pi^0$ through
the light meson chiral Lagrangian.
\bea
+i\half G_A\Tr(\overline{ {\cal{H}}}{}'
v\cdot{{\cal{A}}} {\cal{H}} ) & \rightarrow&
\nonumber \\
& & 
\!\!\!\!\!\!\!\!\!\!\!\!\!\!\!\!\!\!\!\!\!\!\!\!
\!\!\!\!\!\!\!\!\!\!\!\!\!\!\!\!\!\!\!\!\!\!\!\!
\!\!\!\!\!\!\!\!
-\frac{i G_A}{f_\pi}(\overline{D_s}'_\mu D_s^\mu - \overline{D_s}' D_s)
\left(\frac{-2}{\sqrt{3}}\right)v_\nu\partial^\nu \eta^0
\eea
The amplitude for the decay
is therefore:
\be
\frac{\Delta M}{2f_\pi}\delta_{\eta\pi^0}G_A \qquad
\delta_{\eta\pi^0}=\left( \frac{2m_\pi^2(m_u-m_d)}{
(m_\eta^2-m_\pi^2) (m_u+m_d) }\right).
\ee
where $\delta_{\eta\pi^0}$ parameterizes the
$\eta\pi^0$ mixing. $\delta_{\eta\pi^0}$
vanishes in the limit of nonet symmetry,
as a contribution from the $\eta'$ meson will 
exactly cancel the $\eta$ contribution given above. 
Instanton effects, parameterized by the determinant term 
in the $\Sigma$-potential in eq.(6), break the 
nonet symmetry and generate a large contribution 
to the $\eta'$ mass, suppressing the cancellation. 
The singlet axial current anomalies also signal 
a direct coupling of the singlet $\eta'$ to gluons, 
that will modify the nonet coupling of the 
$\eta'$ to hadrons, 
perhaps further suppressing the singlet 
contribution to the mixing with the $\pi^0$. 
In the following we include only the octet 
$\eta$ contribution to the mixing with the 
$\pi^0$. 
Using $\delta_{\eta\pi^0} = (1/2)\times(1/43.7)$ 
\cite{cho}, the widths are given by:
\bea
\Gamma(D_{s}(0^+)\rightarrow 
D_{s}(0^-)+\pi^0)  & &
\nonumber \\
& &\!\!\!\!\!\!\!\!\!\!\!\!\!\!\!\!\!\!\!\!\!\!\!\!
\!\!\!\!\!\!\!\!\!\!\!\!\!\!\!\!\!\!\!\!\!\!\!\!
= (164 \; \makebox{MeV}) \delta_{\eta\pi^0}^2 = 21.5 \;G_A^2 \; 
\makebox{keV}
\eea
\bea
\Gamma(D_{s}(1^+)\rightarrow 
D_{s}(1^-)+\pi^0)  & &
\nonumber \\
& &\!\!\!\!\!\!\!\!\!\!\!\!\!\!\!\!\!\!\!\!\!\!\!\!
\!\!\!\!\!\!\!\!\!\!\!\!\!\!\!\!\!\!\!\!\!\!\!\!
= (164 \; \makebox{MeV}) \delta_{\eta\pi^0}^2 
= 21.5 \;G_A^2 \; \makebox{keV}
\eea

\vskip .1in
\noindent			      
{\em (iv) $B_{s}(0^+,1^+)\rightarrow 			      
B_{s}(0^-,1^-)+\pi^0$ }
\vskip .1in
\noindent

This case is identical to the $D_s$ system discussion
above:
\bea
\Gamma(B_{s}(0^+)\rightarrow 
B_{s}(0^-)+\pi^0)  & &
\nonumber \\
& &\!\!\!\!\!\!\!\!\!\!\!\!\!\!\!\!\!\!\!\!\!\!\!\!
\!\!\!\!\!\!\!\!\!\!\!\!\!\!\!\!\!\!\!\!\!\!\!\!
= (164 \; \makebox{MeV}) \delta_{\eta\pi^0}^2 
= 21.5 \;G_A^2 \; \makebox{keV}
\eea
\bea
\Gamma(B_{s}(1^+)\rightarrow 
B_{s}(1^-)+\pi^0)  & &
\nonumber \\
& &\!\!\!\!\!\!\!\!\!\!\!\!\!\!\!\!\!\!\!\!\!\!\!\!
\!\!\!\!\!\!\!\!\!\!\!\!\!\!\!\!\!\!\!\!\!\!\!\!
= (164 \; \makebox{MeV}) \delta_{\eta\pi^0}^2 
= 21.5 \;G_A^2 \; \makebox{keV}
\eea

\vskip .1in
\noindent
{\em (v) $D_{s}(1^+)\rightarrow 
D_{s}(0^-)+2\pi$ }
\vskip .1in
\noindent

The analysis of the $2\pi$ transitions is more complicated
and involves effects from the $0^+$ nonet. These
effects are not expected to be cleanly separable from
other resonances higher in the spectrum of the light
quark system. Nonetheless, these probably indicate the 
correct order of the effect, and we include them here because they
are predicted consequences of the model.

In the model the decay can proceed via 
decay to a virtual $\tilde{\sigma}$,
that then converts to the $2\pi$ state,
$D_{s}(1^+)\rightarrow 
D_{s}(0^-)+(\sigma\rightarrow 2\pi)$. The relevant componet
of the $\tilde{\sigma}$ field, that is a $3\times 3$ matrix,
is the $(33)$ component. This must then mix with the $(11)$
and $(22)$ components to produce the pions. 

The relevant HL meson operator is the term from eq.(\ref{chiral2})
of the form:
\bea
\frac{g_A}{2f_\pi} \Tr(\overline{ {\cal{H}}}'\gamma^5\gamma_\mu 
(\partial^\mu\tilde{\sigma}-i[{\cal{V}}^\mu,\tilde{\sigma}])
  {\cal{H}}) 
+\;\;\; ...
\eea
In component form this becomes:
\be
-\frac{ig_A}{f_\pi}D'_{s\mu}{}^\dagger\cdot D_s 
(\frac{\sqrt{2}}{\sqrt{3}}\partial^\mu\sigma^0 
- \frac{2}{\sqrt{3}}\partial^\mu\sigma^8)
\ee

The mixing of the $\sigma^0$ and $\sigma^8$ with
$2\pi$ is controlled by the light sector chiral Lagrangian
of eq.(5). Using the replacement, $\Sigma\rightarrow 
\xi\tilde{\sigma}\xi$ eq.(5) becomes:
\be
\frac{1}{4}\Tr(\partial\tilde{\sigma} -i[{\cal{V}}_\mu,\tilde{\sigma}])^2
+
\frac{1}{4}\Tr\{{\cal{A}}_\mu,\tilde{\sigma}\}^2 + ...
\ee
We shift $\sigma^0= \sqrt{3/2}f_\pi + \sigma$ 
and expand the currents to obtain the effective
couplings to the $2\pi$:
\be
\rightarrow \frac{1}{f_\pi\sqrt{3}}[(\partial\vec{\pi})^2 -m_\pi^2(\vec{\pi})^2]
\left[\sqrt{2}\sigma^0 + \sigma^8\right] 
\ee

Putting this together gives the 
$D_{s}(1^+)\rightarrow 
D_{s}(0^-)+ \pi^0\pi^0$ amplitude:
\be
\frac{2g_A}{3f_\pi^2}\epsilon^\mu q_\mu(q^2-4m_\pi^2)
\left[\frac{1}{q^2-m^2_{\sigma^0}} + \frac{1}{q^2-m^2_{\sigma^8}}\right]
\ee
where $q^2 =(p_1 + p_2)^2$ is the $2\pi$ system invariant mass.
The $\pi^+\pi^-$ amplitude is $\sqrt{2}$ larger.

The resulting widths are controlled by
the 
$\Delta M(D_s(1^+)-D_s(0^-)) = 491$ MeV.
The phase space integrals are extremely sensitive to scalar 
masses varying by over an order of magnitude when the lighter 
singlet mass is varied over the range $0.8$ to $1.2$ GeV.   
We will give the values for $m_{\sigma^0} = 1.0$ GeV with a 
heavier octet scalar at $1.5$ GeV (note the singlet-octet 
splitting is opposite for scalars and pseudoscalars).  
As in our discussion of the $\eta$-$\pi^0$ mixing, the coupling 
of the singlet meson to hadrons is also uncertain due to 
the mixing with gluons.   
Nevertheless we give representative widths using 
nonet couplings and the scalar masses given above:
\bea
\Gamma(D_s(1^+)\rightarrow D_s(0^-)\pi^0\pi^0) & = & 
6.4 g_A^2  
=2.3\; \makebox{keV}
\nonumber \\
\Gamma(D_s(1^+)\rightarrow D_s(0^-)\pi^+\pi^-) & = & 
5.4 g_A^2  
= 1.9\; \makebox{keV}
\nonumber \\
\Gamma(D_s(1^+)\rightarrow D_s(0^-)\pi\pi) & = & 
4.2\;\; \makebox{keV}
\eea
where $g_A=0.6$ is used.
The corresponding widths in the $B_s$ system are
significantly smaller due to the reduced phase space:
\bea
\Gamma(B_s(1^+)\rightarrow B_s(0^-)\pi^0\pi^0) & = & 
0.19 g_A^2  
=0.07\; \makebox{keV}
\nonumber \\
\Gamma(B_s(1^+)\rightarrow B_s(0^-)\pi^+\pi^-) & = & 
0.13 g_A^2  
= 0.05\; \makebox{keV}
\nonumber \\
\Gamma(B_s(1^+)\rightarrow B_s(0^-)\pi\pi) & = & 
0.12\; \makebox{keV}
\eea

\vskip .1in
\noindent
{\em (vi) $D_{s}(1^+)\rightarrow 
D_{s}(0^-)+3\pi$ }
\vskip .1in
\noindent

This decay is allowed by phase space, and proceeds
through $\omega-\phi$ mixing. It is highly suppressed
by chiral symmetry, as well as a
strong OZI-rule suppression. We do not
consider it further in the present paper.

\vskip .2in
\noindent
{\bf (B) Intramultiplet Transitions}
\vskip .1in

The chiral structure of the theory controls the 
intramultiplet decays of the 
form $D(1^\pm)\rightarrow D(0^\pm)+\pi$. These decays 
within multiplets proceed through the $g_A$ coupling term:
\be
+\frac{g_A}{2f_\pi} 
\left[  \Tr(\overline{ {\cal{H}}}'\gamma^5\gamma_\mu 
\{ {\cal{A}}^\mu,\tilde{\sigma} \}
  {\cal{H}}') 
 -\Tr(\overline{ {\cal{H}}}{}\gamma^5\gamma_\mu 
 \{ {\cal{A}}^\mu,\tilde{\sigma} \}
 {\cal{H}})
\right]
\ee
with $\tilde{\sigma} \rightarrow f_\pi I_3$.
Such transitions are relevant
only for the charmed mesons. The 
intramultiplet hyperfine
splitting in mass between the $1^\pm$
and $0^\pm$ states is too small in
the $B$ mesons (and even smaller
in the $ccq$, $bcq$ and $bbq$ baryons)
to allow this decay. 

The resulting decay widths are:
\bea
\Gamma(D^{*+}(1^-)\rightarrow D^+(0^-)\pi^0)
 &=&  181 g_A^2 \; \makebox{keV}  = 65.2 \; \makebox{keV} \nonumber \\
\Gamma(D^{*+}(1^-)\rightarrow D^0(0^-)\pi^+)
 &=&  83 g_A^2 \; \makebox{keV} = 30.1 \; \makebox{keV} \nonumber \\
\eea
where $g_A=0.6$ was used.
The identical widths are obtained
for the $1^+\rightarrow 0^+ +\pi$ modes.
\bea
\Gamma(D^{*+}(1^+)\rightarrow D^+(0^+)\pi^0)
&=&  181 g_A^2 \; \makebox{keV}  = 65.2 \; \makebox{keV} \nonumber \\
\Gamma(D^{*+}(1^+)\rightarrow D^0(0^+)\pi^+)
&=& 83 g_A^2 \; \makebox{keV} = 30.1 \; \makebox{keV} \nonumber \\
\eea

\section{Electromagnetic Transitions}

In the static limit, heavy-light mesons can be used to define 
the electromagnetic properties of the tethered constituent--quark.   
In fact, it has sometimes been suggested that the constituent 
quark mass be defined through the meson magnetic moment in this limit.   

The M1 electromagnetic transitions govern the intramultiplet
processes, $1^\pm \rightarrow 0^\pm \gamma$, while the 
E1 transitions
govern intermultiplet processes, $(1^+,0^+)\rightarrow (1^-,0^-)\gamma$.
There are significant finite heavy--quark mass corrections 
particularly for the $D_s$ system.  We observe below 
that the $1^-\rightarrow 0^-\gamma$ M1 transition amplitude,
and the three E1  
transition amplitudes,
$1^+\rightarrow 1^-\gamma$, $1^+\rightarrow 0^-\gamma$
and $0^+\rightarrow 1^-\gamma$, 
 receive a common overall coefficient $r_{\bar{Q}q}$.
We find:
\be
r_{\bar{Q}q} = \left( 1 - \frac{m^*_q e_{\bar{Q}} }{m^*_{\bar{Q}} e_q} \right)
\ee
where $m^*$ and $e$  are the mass and charge of the constituent--quarks. 
In the $D_s$ system the anti--charm quark has a charge of 
$-2/3$ and the strange quark charge $-1/3$ 
leading to a large suppression (see the Appendix of 
\cite{eichten2}):
\be
r_{\bar{c}s}=\left(1-\frac{2m^*_s}{m^*_c} \right)
\ee
The $D_d$ has a somewhat smaller suppression and the $D_u$
an enhancement.   In the $B$-meson system the situation is 
reversed as the $\bar{b}$-quark has charge $+1/3$ although 
the overall effects are much smaller due to the 
larger mass for the $b$-quark.

We use the usual constituent quark potential model to 
estimate the electromagnetic transition rates.   
For the M1 magnetic transitions 
$1^-\rightarrow 0^-\gamma$ the rate is given by:
\be
\label{M1q_eq}
\Gamma_{\text{M1}}(i\rightarrow f\gamma) = \frac{4\alpha}{3}\mu_{\bar{Q}q}^2 k^3 
(2J_f+1)|\langle f|j_0(kr)|i\rangle|^2,
\ee
where the magnetic dipole moment is:
\be
\mu_{\bar{Q}q} = \frac{m_Q^*e_q-m_q^*e_{\overline{Q}}}{2m_Q^*m_q^*}
= \frac{e_q}{2m_q^*}r_{\bar{Q}q}  \;\;\;\;\;
\ee
and $k$ is the photon energy.

The strength of the electric-dipole transitions is governed by the size 
of the radiator and the charges of the constituent--quarks.  The E1 
transition rate is given by
\be
\Gamma_{\text{E1}}(i\rightarrow f+\gamma) = 
\frac{4\alpha <\!e_{\rm avg}\!>^2}{27} 
k^3 (2J_f+1)|\langle f|r|i\rangle |^2 {\cal S}_{if}\;\;\; ,
\end{equation}
where the mean charge is 
\be
<\!e_{\rm avg}\!> = \frac{m_Q^*e_q-m_q^*e_{\overline{Q}}}{m_Q^*+m_q^*}
= \frac{{e_q}m^*_Q r_{\bar{Q}q}}{m^*_Q+m^*_q} \;\;\; ,
\label{E1q_eq}
\ee
$k$ is the photon energy, and the statistical factor, ${\cal S}_{if}$,
for $(i,f) = (0^+,1^-)$ is 1,  for $(1^+,1^-)$ is $2/3$,
and for $(1^+,0^-)$ is 1.

To evaluate the factor $r_{\bar{Q}q}$ we use the constituent--quark masses:
\bea
& & m_u^*=m_d^*=350 \;\; \makebox{MeV} \nonumber \\
& & m_s = 480 \;\; \makebox{MeV} \nonumber \\ 
& & m_c^* = (3M(J/\Psi)-M(\eta_c))/8 = 1530 \;\; \makebox{MeV} \nonumber \\ 
& & m_b^* = (3M(\Upsilon)-M(\eta_b))/8 = 4730 \;\; \makebox{MeV} \nonumber \\
\eea
This in turn leads the $r_{\bar{Q}q}$ factors:
\bea
& & r_{\bar{c}u} = 1.23 \qquad  r_{\bar{b}u} = 0.85  \nonumber \\
& & r_{\bar{c}d} = 0.54 \qquad  r_{\bar{b}d} = 1.07  \nonumber \\
& & r_{\bar{c}s} = 0.38 \qquad  r_{\bar{b}s} = 1.10  
\label{rmq_eq}
\eea 
With
these $r_{\bar{Q}q}$ factors we can see the  large
cancellation between the light $(d,s)$ quark moment and the charm
quark moment.  Using the measured total width of the
$D^{+*}$ to set the pionic transition. The partial rates  
for photonic decays in the $D^{0*}$ and $D^{+*}$ systems can be calculated.
The uncertainty in the total width drops out for the ratio of partial widths: 
\be
\frac{\Gamma[D^{+*}\rightarrow D^+ +\gamma]}
{\Gamma[D^{0*}\rightarrow D^0 +\gamma]} 
= \frac{1.6\pm 0.4}{27.4\pm 2.1}
 = 0.058 \pm 0.015 
\ee 
This implies:
\be
  \left|\frac{\mu(D^{+*})}{\mu(D^{0*})}\right| = 0.24 \pm 0.03\;|_{expt} =  
   \half \left( \frac{r_{\bar{c}d}}{r_{\bar{c}u}} \right)
   = 0.22\;|_{theory}
\ee 
In the HQ limit this ratio is $0.5$. Hence the finite charm quark mass
provides a large cancellation for the $D^{+*}$ system. This suppression
of the M1 transition will be even larger for the $D_s^{+*}$ system as 
$r_{\bar{c}s}< r_{\bar{c}s}$.
Rates for the allowed M1 transitions are given in Table II. 

The same cancellation that appears for the M1 transition is operative 
for the E1 transitions. In the $D_s$ system this greatly suppresses the 
rate for the $(0^+,1^+) \rightarrow (0^-,1^-) + \gamma$ allowed 
E1 transitions. 
The E1 transition rates and photon energies are 
also presented in Table II. 

The observed ratio of branching fractions   
$(D_s(1^-)\rightarrow D_s(0^-)\pi^0)/
\Gamma(D_s(1^-)\rightarrow D_s(0^-)\gamma) = 0.062\pm 0.026$
is large compared to our prediction of $0.018$. This may indicate that
$r_{\bar cs}$ is more suppressed than our estimate in eq.(\ref{rmq_eq}).
If we implement the experimental value for this ratio,
then the E1 radiative transitions of Table II for the $\bar{c}s$ system
should be reduced by a factor of $\sim 3$.

In the $B$-system there is no suppression for the $B_d$
and $B_s$ transitions are slightly enhanced by the $r_{\bar{b}q}$
factors.
There is a small suppression in the $B_u$ states.
The resulting electromagnetic transition rates and photon energies for the 
narrow B states are presented in Table II. 

For the $1^+\rightarrow 0^+\gamma$ M1 transition
we define the coefficient $r'_{\bar{Q}q}$:
\be
r'_{\bar{Q}q} = \left( 1 + 3\frac{m^*_q e_{\bar{Q}} }{m^*_{\bar{Q}} e_q} \right)
\ee
The decay rate is given by:
\be
\label{M1q_eq}
\Gamma_{\text{M1}}(i\rightarrow f\gamma) = 
\frac{4\alpha}{3}\mu^{\prime 2}_{\bar{Q}q} k^3 
(2J_f+1)|\langle f|j_0(kr)|i\rangle|^2,
\ee
where the effective magnetic dipole moment $\mu'_{\bar{Q}q}$ is now:
\be
\mu'_{\bar{Q}q} = \frac{-m_Q^*e_q-3m_q^*e_{\overline{Q}}}{6m_Q^*m_q^*}
= -\frac{e_q}{6m_q^*}r'_{\bar{Q}q}  \;\;\;\;\;
\ee
and $k$ is the photon energy,
\be
r'_{\bar{c}s}=2.88  \qquad   r'_{\bar{b}s}=0.70.
\ee
These decay rates are also included
for the $D_s$ and $B_s$ systems in Table II.

Finally,
we have ignored mixing between the two $1^+$ p-wave
mesons as the parity partner of the s-wave mesons has $j_\ell=1/2$
which does not mix with the $j_\ell=3/2$ state at leading order in
the heavy quark expansion. The total angular momentum
of the light quark, $j_\ell$, is conserved in the heavy quark limit.

\section{Doubly-Heavy Baryons}

We will provide only a schematic discussion of the
corresponding situation in the doubly-heavy baryons,
and defer tabulating detailed results. These systems 
provide interesting targets of opportunity in the spectroscopy 
of QCD, but are challenging to reconstruct. For
some recent reviews and relevant information see \cite{double}.
A chiral constituent--quark model similar to \cite{Bardeen}, has been
developed as well for these systems \cite{ebert}.

There are four distinct doubly-heavy baryon systems, 
each transforming as flavor $SU(3)$ triplets,
$[cc]_{J=1}(u,d,s)$, $[bc]_{J=0}(u,d,s)$,
$[bc]_{J=1}(u,d,s)$, and $[bb]_{J=1}(u,d,s)$.
There is evidence for doubly-charm baryons
in the SELEX data \cite{cooper}. 

These systems are interesting because of heavy
quark symmetry, since the $[QQ]$ subsystem has a large
mass, of order $\sim 2m_{Q}$, and forms, in
the subsystem groundstate, a tightly bound
anti-color triplet combination,
that can be viewed as a heavy spin-1 or spin-0 antiquark, 
$[QQ]\sim \overline{Q}'$. 
Hence, doubly-heavy baryons can be viewed as
ultraheavy mesons, $[QQ]q \sim \overline{Q}' q$.
The hyperfine mass splittings in these systems are suppressed. 

The doubly-heavy baryon groundstates of the form
$[QQ]_{J=1} q$ will consist of multiplets
containing $(1/2^+, 3/2^+)$ heavy spin fields.
For example, $[cc](u,d)$ groundstate 
will contain one $I=\half$, spin-$1/2$ baryon, and
one $I=\half$ spin-$3/2$ baryon (this is
the analogue in the $[ss](u,d)$ of a multiplet
containing the $(\Xi^0,\Xi^-)$ spin-$1/2$ baryon
from the octet, and the $(\Xi^{*0},\Xi^{*-})$ spin-$3/2$ 
resonance from the decouplet).
The hyperfine splitting
mass differences within the multiplets,
between the spin-$3/2$ and spin-$1/2$ members,
have been estimated in \cite{ebert}.
Note that in the case of the $[cb]_{J=1}$
we have the normal $(1/2^+, 3/2^+)$ multiplet, while
in the $[cb]_{J=0}$ system the 
spin--$3/2$ partner is absent. 

The parity partner states are correspondingly
$p$-wave resonances $(1/2^-, 3/2^-)$. 
The intramultiplet mass splitting will be 
approximately identical to
the case of the groundstate.
The intramultiplet
chiral mass gap, for each system is therefore
given by $\Delta M(\infty)= 349 \pm 35$ MeV.

The four systems with a strange quark will display
narrow resonances in analogy to the $D_s(0^+,1^+)$.
These will have blocked kaonic decays to the groundstate.
They will decay mesonically
in a manner identical to the HL meson system
with the correspondence $(0^+,1^+)\leftrightarrow 
((1/2)^-, (3/2)^-)$. All of our computed intramultiplet
widths will correspond identically, but
will have significant
modifications due to the hyperfine splittings affecting
the phase space and kinematic factors. 

The electromagnetic transitions will correspond 
with the meson case in
a likewise fashion.  The intramultiplet
transitions will be suppressed because of
the reduced phase space.
The widths in the $[bb]s$
system will have suppressed cancellations and will be
predominantly governed by the light--quark terms alone.

\section{Conclusions}

We have examined the chiral structure of the HL meson
system in QCD.
The groundstate $(0^-,1^-)$ multiplet is paired 
with the $(0^+,1^+)$
multiplet through chiral symmetry. 
The physical significance of this statement is
that the linear combinations of these states,
${\cal{H}}_{L}$ and
${\cal{H}}_{R}$ form the definite
representations under $SU(3)_L\times SU(3)_R$
of $(3,1)$ and $(1,3)$ respectively. 
 
The spontaneously breaking of chiral symmetry in
QCD leads to a mass term that elevates the $(0^+,1^+)$
above the $(0^-,1^-)$ by an amount $\Delta M$.
Chiral invariance implies that $\Delta M$
satisfies the Goldberger-Treiman relation, $\Delta M = g_\pi f_\pi$.
This value of $g_\pi$ is essentially the pure pionic coupling
of a constituent chiral quark, such as found in, \eg, the
Georgi-Manohar model \cite{manohar}. In a nucleon, a simple constituent--quark
picture would predict $g_{NN\pi} \approx 3 g_\pi$. 
The analogue of $\Delta M$ for the nucleon system is the nucleon mass,
$m_N$, that satisfies the traditional Goldberger-Treiman relation,
$m_N = g_{NN\pi} f_\pi$. $\Delta M \approx m_N/3$ is
close to its observed value in the BABAR data of $349$ MeV. 
In a remarkable sense, the HL meson is displaying the chiral 
dynamical properties
of a single light--quark, that is ``tethered'' to the heavy
quark.

Our hypothesis fits the recent observation of the narrow resonance in
$D_s\pi^0$ seen by the BABAR collaboration. 
In general, 
$D$ mesons form $SU(3)$ flavor triplets,
and approximate heavy spin multiplets.  The
nonstrange $I=\half$ $(0^+,1^+)$
states can undergo $I=1$ transitions to the $I=\half$ $(0^-,1^-)$
by emission of a single pion. The coupling strength of this transition
is governed by the GT relation, and these states are therefore broad,
with total widths predicted to be $490$ MeV.

The $D_s(0^+,1^+)$ resonance, on the other hand, is kinematically 
forbidden from undergoing its principal decay transition  
$D_s(0^+,1^+) \rightarrow D(0^-,1^-)+K$. The observed decay
$D_s(0^+) \rightarrow D_s(0^-)+\pi^0$ violates isospin, and
therefore requires $SU(3)$ breaking effects. In the present paper
we have addressed the main  kinematically allowed effects
$D_s(0^+) \rightarrow D_s(0^-)+\pi^0$, and 
$D_s(1^+) \rightarrow D_s(0^-)+2\pi$. The widths are small,
consistent with the narrowness of the observed systems.

We have also tabulated the electromagnetic transitions.
Due to the cancellations between light--quark and heavy--quark
amplitudes, the intermultiplet E1
rates for the $\bar{c}s$ system are suppressed. This explains why
the photonic decays are not seen in the BABAR data. 
In the analogue $\bar{b}s$ system the rates are not
suppressed by a similar cancellation, 
and the electromagnetic transition widths are significantly
larger.

It is important to realize that these heavy--quark
and chiral symmetry arguments are 
quite general. What has been discovered by BABAR is a {\em phenomenon}.
Analogue effects will be seen in the $B$ meson system
as well as the doubly-heavy baryons. 

In the $B_s$ system we expect a splitting between the $(0^-,1^-)$
$SU(3)$ triplet groundstate mesons and the analogue $(0^+,1^+)$
resonance multiplet with a mass of 
$349\pm {\cal{O}}(\Lambda_{QCD}/m_{charm})$
or, $\sim 349 \pm 35$ MeV.  Again, the channel 
$B_s(0^+,1^+) \rightarrow B_s(0^-,1^-)+K$ will be kinematically
blocked, while the $B_{u,d}(0^+,1^+)$ states will have similar narrow 
meson-transition widths given.
We have also described schematically how the doubly--heavy 
baryons $[QQ]q$ will present  an analogous 
situation.

Heavy-light meson states may be classified according to
the total angular momentum carried by the light quark to
leading order in the heavy quark limit.   In the present paper
we have focused on the $j_\ell = 1/2$ parity doubled supermultiplet
combining the s-wave $(0^-,1^-)$ mesons and the
p-wave $(0^+,1^+)$ mesons.   As mentioned in the introduction,
we expect all heavy-light states to be classified into parity
doubled supermultiplets where the mass splitting between
parity partners is governed by the Goldberger-Treiman
relation.   For example, the $j_\ell = 3/2$ supermultiplet
consists of the p-wave $(1^+,2^+)$ mesons with $j_\ell=\ell+1/2$
and the d-wave $(1^-,2^-)$ mesons with $j_\ell=\ell-1/2$, similarly
for the higher angular momentum states.   In QCD, string
models and linear potential models, the meson states are
expected to be identified with linear Regge trajectories having
a common slope \cite{dip}.   A remarkable
consequence of this observation is that the mass splitting
between parity partners for the higher angular momentum
supermultiplets will remain constant, ie. the Yukawa coupling
constant, $g_\pi$, governing the left-right transitions
in eq.(12) will be universal.   The dynamical breaking of the
light quark chiral symmetries in QCD is apparently associated
with the observed shifts in the opposite parity Regge trajectories
by about half a unit of the Regge spacing from a completely
parity doubled picture of the Regge trajectories.

In a larger sense, parity doubling, which is required
in any dynamics that can putatively
restore spontaneously broken chiral symmetry of QCD
without upsetting confinement, 
appears to play an important role in the real world,
and is controlling the chiral physics of HL systems.
\vskip .1in
\noindent
{\bf Acknowledgements}
\vskip .1in
\noindent
We thank J. Butler, C. Quigg and S. Stone 
for discussions.
Research supported by the U.S.~Department of Energy  
grant DE-AC02-76CHO3000.  

\vskip .5in
\noindent
{\bf Appendix A: Normalization Conventions }
\vskip .1in

Consider a complex scalar field, $\Phi$, with the Lagrangian:
\beq
\partial_\mu\Phi^\dagger \partial^\mu\Phi - (M+\delta M)^2
\Phi^\dagger \Phi
\eeq
Define $\Phi' = \sqrt{2M}\exp(iMv\cdot x)\Phi$ ($\Phi'$ destroys
incoming momentum $Mv_\mu + p_\mu$)
and the Lagrangian becomes to order $1/M$:
\beq
 iv_\mu\Phi'{}^\dagger \partial^\mu\Phi' - \delta M
\Phi'{}^\dagger \Phi'
\eeq
Now let ${\cal{H}}_v =
\half(1-\slash{v})i\gamma^5\Phi'$ and write in terms of traces
(the field ${\cal{H}}_v$ with these conventions annihilates 
an incoming meson state $\ket{B}$):
\beq
-i\half \Tr(\overline{\cal{H}}_v v\cdot \partial {\cal{H}}) 
+ \delta M\half \Tr(\overline{\cal{H}} {\cal{H}}_v )
\eeq
Thus, when the Lagrangian is
written in terms of ${\cal{H}}$ and ${\cal{H}}'$
the normal sign conventions are those of the vector mesons,
and opposite those of scalars, \ie,
the term in the Lagrangian $+\half \delta M \Tr(\overline{\cal{H}}H)$  
an {\em increase} in the
${\cal{H}}$ multiplet mass by an amount $ \delta M$. 
A properly normalized kinetic term is:  
$\half \Tr(\overline{{\cal{H}}} i\slash{\partial}
{\cal{H}} )
=
-i\half\Tr({\cal\overline{{\cal{H}}}}v\cdot \partial{\cal{H}})$.


\begin{table*}
\begin{center}
\caption{The heavy-light spectrum compared to experiment. 
We report the difference between the excited 
state masses and the ground state ($D$ or $B$) in each case.
We have assumed that $\Delta M(m_c)= \Delta M(m_b) =
\Delta M(\infty) =349$ MeV.}  
\begin{tabular}{llllll} 
\hline
\hline
\multicolumn{3}{c}{charmed meson masses [MeV]} 
& \multicolumn{3}{c}{bottom meson masses [MeV]} \\
\cline{1-3}\cline{4-6}
               & model          & experiment                      & 
               & model          & experiment                      \\
\hline
\hline
$D^{*0}-D^0$              & 142~[a]        & $142.12 \pm 0.07$       & 
$B^{*0}-B^0$              & 46~[a]         & $45.78  \pm 0.35$         \\ 
$D^{*+}-D^+$              & 141~[a]        & $140.64 \pm 0.10$       & 
$B^{*+}-B^+$              & 46~[a]         & $45.78  \pm 0.35$         \\
$D_s^{*+}-D_s^+$          & 144~[a]        & $143.8  \pm 0.41$       & 
$B_s^{*+}-B_s^+$          & 47~[a]         & $47.0   \pm 2.6$          \\
\hline
$D^0(0^+)-D^0$            & 349            &                         & 
$B^0(0^+)-B^0$            & 349            &                           \\
$D^+(0^+)^-D^+$           & 349            &                         & 
$B^+(0^+)^-B^+$           & 349            &                           \\ 
$D_s^+(0^+)^-D_s^+$       & 349~[a]        & $349 \pm 1.3$~[b]       & 
$B_s^+(0^+)^-B_s^+$       & 349            &                           \\ 
\hline
$D^0(1^+)-D^0(0^+)$       & 142            &                         & 
$B^0(1^+)-B^0(0^+)$       & 46             &                           \\
$D^+(1^+)^-D^+(0^+)$      & 141            &                         & 
$B^+(1^+)^-B^+(0^+)$      & 46             &                           \\ 
$D_s^+(1^+)^-D_s^+(0^+)$  & 144            &                         & 
$B_s^+(1^+)^-B_s^+(0^+)$  & 47             &                           \\ 
\hline
\hline
\end{tabular}
\end{center}
\label{table_exp}
[a] Experimental input to model parameters fit. 
  [b] BaBar result \cite{BABAR}. \newline
\end{table*}

\begin{table*}
\label{widths}
\begin{center}
\caption{The predicted hadronic and electromagnetic transistion rates for
narrow  $j_l^P = 1/2^-(1S)$ and $ j_l^P = 1/2^+(1P)$ heavy-light states.
``Overlap'' is the reduced matrix element overlap integral; ``dependence''
refers to the sensitive model parameters, as defined in the text. We
take $G_A=1$ and extract $g_A$ from a fit to the $D^{+*}$ total width.
Note that the $\bar{c}s$
transitions are sensitive to $r_{\bar{c}s}$;
if we implement the observed ratio of branching fractions   
$(D_s(1^-)\rightarrow D_s(0^-)\pi^0)/
\Gamma(D_s(1^-)\rightarrow D_s(0^-)\gamma) = 0.062\pm 0.026$
then the E1 radiative transitions for the $\bar{c}s$ system
should be reduced by a factor of $\sim 3$
}
\begin{tabular}{cllllll} 
\hline
\hline
system &  transition & Q(keV) & overlap & dependence & $\Gamma$ (keV) & exptl BR \\
\hline
\hline
$(c\bar u)$ &  $1^- \rightarrow 0^- + \gamma$ & $137$ &  $0.991$ & 
                              $r_{\bar{c}u}$ & $33.5$ & ($38.1\pm 2.9$)\% \\
            &  $1^- \rightarrow 0^- + \pi^0$  & $137$ &          &
                                       $g_A$ & $43.6$ & ($61.9\pm 2.9$)\% \\
            &  ~~~~~total                     &       &          & 
                                             & $77.1$ &          \\
\hline
$(c\bar d)$ &  $1^- \rightarrow 0^- + \gamma$ & $136$ &  $0.991$ & 
                              $r_{\bar{c}d}$ & $1.63$ & ($1.6\pm 0.4$)\%  \\
            &  $1^- \rightarrow 0^- + \pi^0$  & $38$  &          &  
                                       $g_A$ & $30.1$ & ($30.7\pm 0.5$)\% \\
            &  $1^- \rightarrow 0^- + \pi^+$  & $39$  &          &
                                       $g_A$ & $65.1$ & ($67.7\pm 0.5$)\% \\
            &  ~~~~~total                     &       &          & 
                                             & $96.8$ & $96\pm 22$        \\
\hline
$(c\bar s)$ &  $1^- \rightarrow 0^- + \gamma$ & $138$ & $0.992$  &
                              $r_{\bar{c}s}$ & $0.43$ & ($94.2\pm 2.5$)\% \\ 
            &  $1^- \rightarrow 0^- + \pi^0$  & $48$  &          &
                   $g_A \delta_{\eta\pi0}$ & $0.0079$ & ($5.8\pm 2.5$)\%  \\
            &  ~~~~~total                     &       &          & 
                                             & $0.44$ &          \\
\hline
$(c\bar s)$ &  $0^+ \rightarrow 1^- + \gamma$ & $212$ & $2.794$  & 
                               $r_{\bar{c}s}$ & $1.74$ &          \\
            &  $0^+ \rightarrow 0^- + \pi^0$  & $297$ &          &
                     $G_A \delta_{\eta\pi0}$ & $21.5$ &          \\
            &  ~~~~~total                     &       &          & 
                                             & $23.2$ &          \\
\hline
$(c\bar s)$ &  $1^+ \rightarrow 0^+ + \gamma$ & $138$ & $0.992$  & 
                               $r'_{\bar{c}s}$ & $2.74$ &          \\
	    &  $1^+ \rightarrow 0^+ + \pi^0$  & $48$  &          &
                   $g_A \delta_{\eta\pi0}$ & $0.0079$ &          \\
            &  $1^+ \rightarrow 1^- + \gamma$ & $323$ & $2.638$  & 
                               $r_{\bar{c}s}$ & $4.66$ &          \\
            &  $1^+ \rightarrow 0^- + \gamma$ & $442$ & $2.437$  & 
                              $r_{\bar{c}s}$ & $5.08$ &          \\
            &  $1^+ \rightarrow 1^- + \pi^0$  & $298$ &          & 
                     $G_A \delta_{\eta\pi0}$ & $21.5$ &          \\
            &  $1^+ \rightarrow 0^- + 2\pi$   & $221$ &          & 
             $g_A \delta_{\sigma_1\sigma_3}$ & $4.2$  &          \\
            &  ~~~~~total                     &       &          & 
                                             & $38.2$ &          \\
\hline
$(b\bar u)$ &  $1^- \rightarrow 0^- + \gamma$ & $46$  & $0.998$  & 
                               $r_{\bar{b}u}$ & $0.78$ &          \\
            &  ~~~~~total                     &       &          & 
                                             & $0.78$ &          \\
\hline
$(b\bar d)$ &  $1^- \rightarrow 0^- + \gamma$ & $46$ &  $0.998$  & 
                                $r_{\bar{b}d}$ & $0.24$ &          \\
            &  ~~~~~total                     &       &          & 
                                             & $0.24$ &          \\
\hline
$(b\bar s)$ &  $1^- \rightarrow 0^- + \gamma$ & $47$ &  $0.998$  &
                              $r_{\bar{b}s}$ & $0.15$ &          \\
            &  ~~~~~total                     &       &          & 
                                             & $0.15$ &          \\
\hline
$(b\bar s)$ &  $0^+ \rightarrow 1^- + \gamma$ & $293$ & $2.536$  & 
                              $r_{\bar{b}s}$ & $58.3$ &          \\
            &  $0^+ \rightarrow 0^- + \pi^0$  & $297$ &          & 
                     $G_A \delta_{\eta\pi0}$ & $21.5$ &          \\
            &  ~~~~~total                     &       &          &
                                             & $79.8$ &          \\
\hline
$(b\bar s)$ &  $1^+ \rightarrow 0^+ + \gamma$ & $47$ & $0.998$  &
                              $r'_{\bar{b}s}$ & $0.061$ &          \\
            &  $1^+ \rightarrow 1^- + \gamma$ & $335$ & $2.483$  & 
                              $r_{\bar{b}s}$ & $56.9$ &          \\
            &  $1^+ \rightarrow 0^- + \gamma$ & $381$ & $2.423$  &
                              $r_{\bar{b}s}$ & $39.1$ &          \\
            &  $1^+ \rightarrow 1^- + \pi^0$  & $298$ &          & 
                     $G_A \delta_{\eta\pi0}$ & $21.5$ &          \\
            &  $1^+ \rightarrow 0^- + 2\pi$   & $125$ &          & 
             $g_A \delta_{\sigma_1\sigma_3}$ & $0.12$ &          \\
            &  ~~~~~total                     &       &          & 
                                            & $117.7$ &          \\
\hline
\hline
\end{tabular}
\end{center}
\label{tbl:width_exp}
\end{table*}

\end{document}